\documentclass[prd,twocolumn,groupedaddress,showpacs,nofootinbib]{revtex4}
\usepackage{graphicx}
\usepackage{dcolumn}
\usepackage{amssymb}
\usepackage{mathrsfs}
\usepackage{amsmath}
\usepackage{epsfig}
\usepackage[dvips]{color}
\usepackage{hhline}

\begin{document}

\title{Neutrinoless double beta decay without Majorana neutrinos}

\author{Pei-Hong Gu}

\email{phgu@seu.edu.cn}

\affiliation{School of Physics, Jiulonghu Campus, Southeast University, Nanjing 211189, China}

\begin{abstract}

It is firmly believed that a signal of neutrinoless double beta decay can unquestionably confirm the Majorana nature of neutrinos. However we notice that a Majorana neutrino mass induced after some neutrinoless double beta decay processes can be accidentally cancelled by another Majorana neutrino mass induced before any neutrinoless double beta decay processes. This realistic cancellation can simultaneously allow an observable neutrinoless double beta decay and a vanishing Majorana neutrino mass. In consequence a future discovery of neutrinoless double beta decay cannot fully rule out the possibility of Dirac neutrinos.

\end{abstract}


\maketitle

After Pauli proposed the idea of neutrino in 1930 and then Fermi achieved the theory of nuclear beta decay in 1934, Racah described the lepton-number-violation process of neutrinoless double beta decay in 1937 \cite{racah1937}. In the neutrinoless double beta decay process, a nucleus containing $N$ neutrons and $Z$ protons decays to a lighter nucleus, changing the nuclear charge by two units while emitting two electrons, i.e. $(N,Z)\rightarrow (N-2,Z+2)+ e^{-}_{}+e^{-}_{}$. From the nuclear beta decay theory, the neutrinoless double beta decay can happen unless the neutrino is assumed to be a Majorana particle \cite{majorana1937}, i.e. the neutrino $\nu$ is identical to the antineutrino $\bar{\nu}$ \cite{racah1937,furry1939}. This is because a Majorana neutrino emitted in the beta decay of one neutron, i.e. $(N,Z)\rightarrow (N-1,Z+1)+ e^{-}_{}+\bar{\nu}_e^{}$, can be reabsorbed in a second one, i.e. $\nu_e^{}(=\bar{\nu}_{e}^{}) + (N-1,Z+1)\rightarrow (N-2,Z+2)+ e^{-}_{}$. The intermediate beta decay in this process is virtual and is forbidden by the nuclear pairing force for some candidate isotopes in neutrinoless double beta decay experiments.

Due to the surprising discovery of parity nonconservation in weak interactions, people realized that the neutrinos emitted and absorbed in weak interactions should be distinguished by their handedness. The neutrinoless double beta decay thus would be exactly forbidden if the Majorana neutrinos were massless. Since 1998 the precise measurements on neutrino oscillations have established a fact that three flavours of neutrinos should be massive and mixing \cite{fukuda1998,fukuda2001,ahmad2001,an2012}. However, the Majorana or Dirac nature of neutrinos remains unknown. In the case the neutrinos are the Majorana particles, their masses should be induced by certain lepton-number-violation interactions such as the famous seesaw mechanisms \cite{minkowski1977,yanagida1979,grs1979,ms1980,mw1980,sv1980,cl1980,lsw1981,ms1981,flhj1989} and the other radiative mechanisms \cite{cl1980,zee1980,zee1986,babu1988,bl2001,knt2003,ma2006}. Alternatively, the masses of Dirac neutrinos can be understood via similar mechanisms with additional symmetry \cite{rw1983,rs1984}.

Besides the original neutrinoless double beta decay process mediated by the Majorana neutrino mass, people explored other lepton-number-violation interactions involving the singly and doubly charged scalars \cite{cl1980,sv1982}, the SUSY particles \cite{mohapatra1986,hkk1995}, the diquark and leptoquark scalars \cite{gu2012}, the diquark and dilepton scalars \cite{bm2002} and so on \cite{gj2008,bhow2013} to realize some alternative neutrinoless double beta decay processes before the Majorana neutrino mass generation. In association with the standard model charged current interactions, any alternative neutrinoless double beta decay processes will eventually lead to a Majorana neutrino mass. This is the so-called black-box theorem \cite{sv1982}. Currently people firmly believe that an observed neutrinoless double beta decay process can confirm not only the violation of lepton number but also the existence of Majorana neutrinos \cite{vergados2002,aee2008,dpr2019}.

Here we courageously consider that a Majorana neutrino mass induced after some neutrinoless double beta decay processes can be accidentally cancelled by another Majorana neutrino mass induced before any neutrinoless double beta decay processes. This realistic cancellation means that an observed neutrinoless double beta decay process can still open a window for a vanished Majorana neutrino mass. In other words, a future discovery of neutrinoless double beta decay cannot completely exclude the possibility of Dirac neutrinos although it apparently establishes the evidence for lepton number violation.

\begin{figure*}
\vspace{6.5cm} \epsfig{file=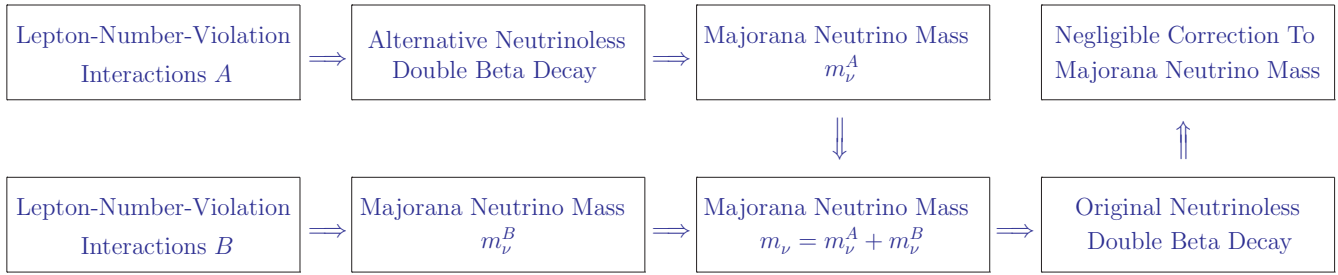, bbllx=11.25cm, bblly=6.0cm,
bburx=21.25cm, bbury=16cm, width=6.5cm, height=6.5cm, angle=0,
clip=0} \vspace{-8.5cm} \caption{\label{pathway} The correlation between Majorana neutrino mass and neutrinoless double beta decay.}
\end{figure*}

\begin{figure*}
\vspace{6cm} \epsfig{file=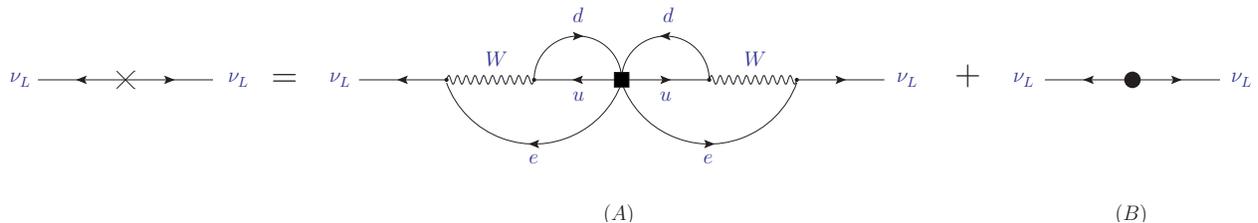, bbllx=12.5cm, bblly=6.0cm,
bburx=22.5cm, bbury=16cm, width=5.5cm, height=5.5cm, angle=0,
clip=0} \vspace{-7.25cm} \caption{\label{numass} The Majorana neutrino mass generation. The black box in the diagram $A$ describes the lepton-number-violation interactions for some alternative neutrinoless double beta decay processes while the black circle in the diagram $B$ includes all the other lepton-number-violation interactions for generating a Majorana neutrino mass. The complete Majorana neutrino mass is the sum of the contributions from the diagrams $A$ and $B$.}
\end{figure*}

\begin{figure}
\vspace{6cm} \epsfig{file=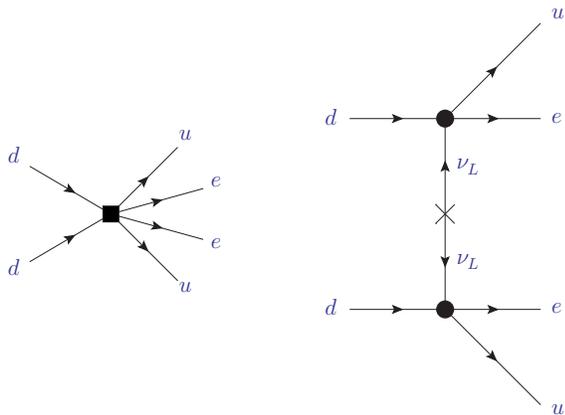, bbllx=5cm, bblly=6.0cm,
bburx=15cm, bbury=16cm, width=6cm, height=6cm, angle=0,
clip=0} \vspace{-5.75cm} \caption{\label{nuless} The alternative and original neutrinoless double beta decay processes at quark level. The black box in the left diagram describes the lepton-number-violation interactions for some alternative neutrinoless double beta decay processes. The right diagram, where the black circles denote the standard model charged current interactions and certain new physics, is the original neutrinoless double beta decay process mediated by a complete Majorana neutrino mass which has included the contribution from the alternative neutrinoless double beta decay processes in the left diagram. }
\end{figure}

\begin{figure}
\vspace{5.75cm} \epsfig{file=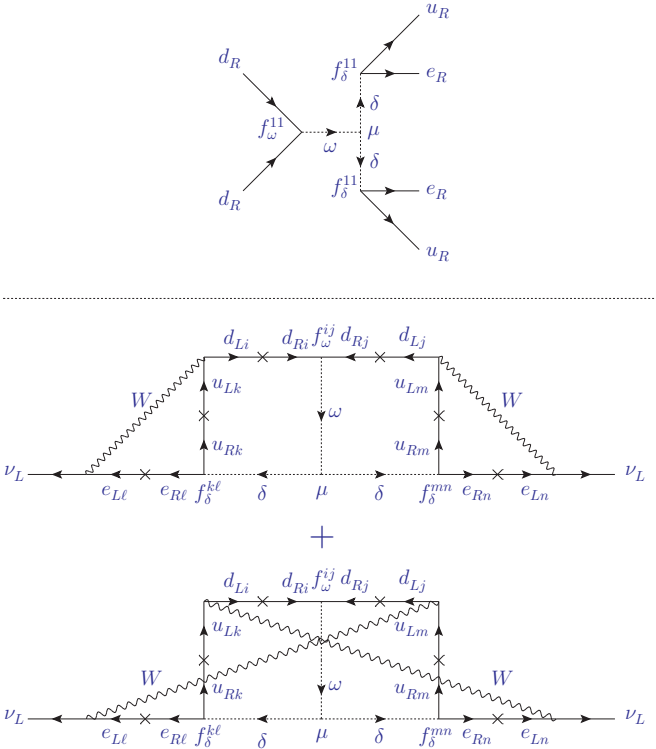, bbllx=5.5cm, bblly=6.0cm,
bburx=15.5cm, bbury=16cm, width=5.5cm, height=5.5cm, angle=0,
clip=0} \vspace{-0.5cm} \caption{\label{right-current} The alternative neutrinoless double beta decay process (upper diagram) and the Majorana neutrino mass (lower diagrams) mediated by the diquark and leptoquark scalars coupling to the standard model right-handed fermions. Here $\omega$ is a color-sextet diquark, $\delta$ is a color-triplet leptoquark, while $f_\omega^{}$, $f_{\delta}^{}$ and $\mu$ denote their Yukawa and cubic couplings. If the lower diagrams only involve the first-generation fermions, they first realize an alternative neutrinoless double beta decay process in the upper diagram and then leave a Majorana neutrino mass according to the black-box theorem. Otherwise, they give a Majorana neutrino mass before any neutrinoless double beta decay processes.}
\end{figure}

\begin{figure}
\vspace{5.5cm} \epsfig{file=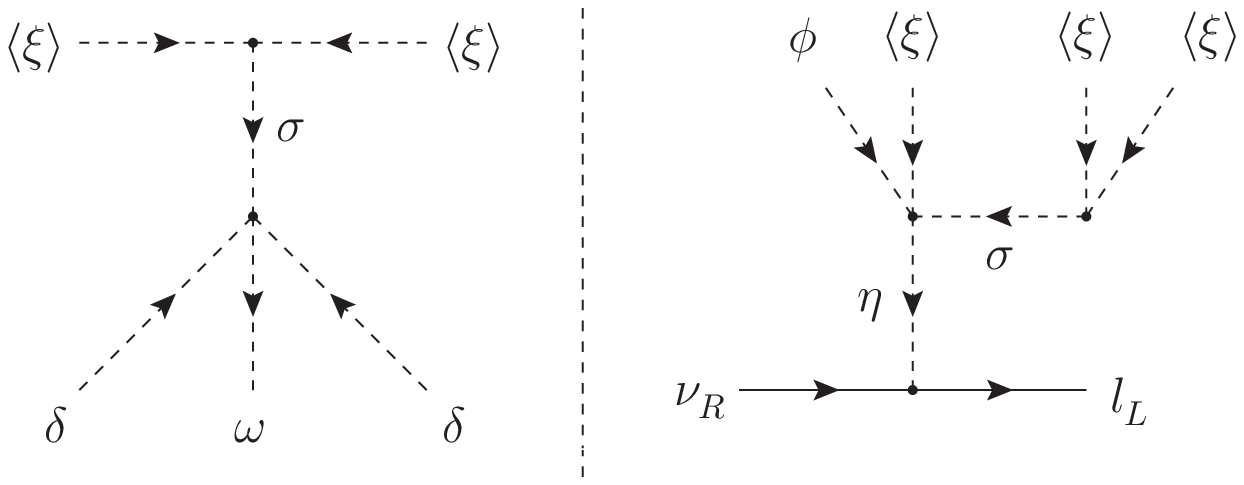, bbllx=5cm, bblly=6.0cm,
bburx=15cm, bbury=16cm, width=5.5cm, height=5.5cm, angle=0,
clip=0} \vspace{-7.75cm} \caption{\label{dirac} The $U(1)_{B-L}^{}$ gauge symmetry breaking for simultaneously generating the lepton-number-violation coupling for neutrinoless double beta decay and the tiny masses for Dirac neutrinos. Here $\omega$ and $\delta$ are the diquark and leptoquark scalars in Fig. \ref{right-current}, $\phi$ and $l_L^{}$ are the standard model Higgs and lepton doublets, while $\xi$ and $\sigma$ are two Higgs singlets, $\eta$ is a new Higgs doublet, $\nu_{R}^{}$ are two right-handed neutrinos. We assign the $B-L$ numbers $(0,+2/3,-2/3,+1,+2,+3,-4)$ for the fields $( \phi, \omega,\delta,\xi,\sigma,\eta,\nu_R^{})$ as the standard model quarks and leptons carry the $B-L$ numbers $+1/3$ and $-1$ as usual. When the Higgs singlet $\xi$ develops its vacuum expectation value, the other Higgs singlet $\sigma$ can be expected to acquire an induced vacuum expectation value. The diquark and leptoquark scalars $\omega$ and $\delta$ then can have a cubic coupling to violate the lepton number. Meanwhile, the new Higgs doublet $\eta$ can obtain a small mixing with the standard model Higgs doublet $\phi$ and then can acquire a small vacuum expectation value after the electroweak symmetry breaking. The Dirac neutrinos thus can obtain their tiny masses in a natural way. Note a new right-handed fermion with the $B-L$ number $+5$ should be necessary to cancel the $U(1)_{B-L}^{}$ gauge anomaly. This fermion which now is massless can decouple above the QCD scale as well as the two right-handed neutrinos to safely avoid the BBN constraint if the $U(1)_{B-L}^{}$ breaking scale is high enough. }
\end{figure}

We conclude the correlation between Majorana neutrino mass and neutrinoless double beta decay in Fig. \ref{pathway}. The related lepton-number-violation interactions are classified in two types: the $A$-type interactions first realize the alternative neutrinoless double beta decay processes and then give a Majorana neutrino mass $m_\nu^A$, while the $B$-type interactions first generate a Majorana neutrino mass $m_\nu^B$ contributing to the complete Majorana neutrino mass $m_\nu^{}=m_\nu^A+m_\nu^B$ and then participate in the original neutrinoless double beta decay process. Clearly the original neutrinoless double beta decay process can give a correction to the Majorana neutrino mass through the black-box theorem \cite{sv1982}. However, this correction is much smaller than the Majorana neutrino mass $m_\nu^{}$ and hence is negligible.

We also show the Majorana neutrino mass generation in Fig. \ref{numass}. The black box in the diagram $A$ describes the $A$-type lepton-number-violation interactions for realizing some alternative neutrinoless double beta decay processes while the black circle in the diagram $B$ includes the $B$-type lepton-number-violation interactions for generating a Majorana neutrino mass. Actually the diagram $A$ is just the black-box theorem while the diagram $B$ comes from the seesaw mechanisms and/or the other radiative mechanisms. We then illustrate the original and alternative neutrinoless double beta decay processes in Fig. \ref{nuless}, where the left diagram describes the alternative neutrinoless double beta decay processes while the right diagram is the original neutrinoless double beta decay process. We should keep in mind that the Majorana neutrino mass in the original neutrinoless double beta decay process has included the contribution from the alternative neutrinoless double beta decay processes in the left diagram.

It is straightforward to see that the original neutrinoless double beta decay process would not occur if the complete Majorana neutrino mass $m_\nu^{}=m_\nu^A+m_\nu^B$ arrived at a zero value. This can be achieved in two cases: (i) $m_\nu^A$ and $m_\nu^B$ are both zero; (ii) $m_\nu^A$ and $m_\nu^B$ are nonzero and opposite. The first case with only one alternative neutrinoless double beta decay process should be trivial because it also implies the absence of any alternative neutrinoless double beta decay processes according to the black-box theorem. This case may be revived by more different alternative neutrinoless double beta decay processes \cite{gj2008,bhow2013} if the related models protest the Majorana neutrino masses induced before the original neutrinoless double beta decay process. As for the second case, it is easy to accommodate the presence of some alternative neutrinoless double beta decay processes \cite{gj2008,bhow2013}. Therefore, we can guarantee a scenario where the neutrinoless double beta decay keeps available while the Majorana neutrino mass becomes vanished. Needless to say, the cancellation in this scenario is contrived from a theoretical point of view. However, such fine-tuning is experimentally acceptable and has to take seriously. This means that the neutrinoless double beta decay experiment cannot perfectly prove the Majorana nature of neutrinos, in contrast to our conviction since 1937.

The above scenario can widely work in numerous combined models of Majorana neutrino mass generation \cite{minkowski1977,yanagida1979,grs1979,ms1980,mw1980,sv1980,cl1980,lsw1981,ms1981,flhj1989,zee1980,zee1986,babu1988,bl2001,knt2003,ma2006} and alternative neutrinoless double beta decay \cite{sv1982,mohapatra1986,hkk1995,gu2012,bm2002,gj2008,bhow2013}. As an example, we demonstrate a specific model in Fig. \ref{right-current}, where a color-sextet diquark and a color-triplet leptoquark only have the Yukawa couplings to the standard model right-handed fermions \cite{gu2012}. The Yukawa interactions only involving the first generation lead to an alternative neutrinoless double beta decay process at tree level and then generate a Majorana neutrino mass at four-loop order, i.e. $m_\nu^B\propto  (f_{\omega}^{11} m_{d}^2)  (f_\delta^{11} m_{u}^{} m_{e}^{})^2_{}$, while the other combinations of Yukawa couplings induce a Majorana neutrino mass at four-loop order before any neutrinoless double beta decay processes, i.e. $m_\nu^A\propto \sum_{ijk\ell mn}^{}( f_{\omega}^{ij} m_{d_i}^{} m_{d_j}^{})(f_{\delta}^{k\ell} m_{u_k}^{} m_{e_\ell}^{})(f_{\delta}^{mn} m_{u_m}^{} m_{e_n}^{} )$, $(i,j,k,\ell,m,n)\neq(1,1,1,1,1,1)$ and then participate in the original neutrinoless double beta decay process. When the two colored scalars are not far above the TeV scale, we can choose the related Yukawa couplings to see the alternative neutrinoless double beta decay process in the ongoing and planned experiments \cite{dpr2019}. With these inputs, the other Yukawa couplings have the flexibility to enforce the complete Majorana neutrino mass vanished, i.e $m_\nu^{} = m_\nu^A+m_\nu^B=0$. Of course such cancellation cannot keep to all orders of perturbation theory. However, we think that the physical values of the Yukawa couplings $f_{\omega,\delta}^{}$ and then the Majorana neutrino masses $m_\nu^{A,B}$ should have included all possible radiative corrections. Now the neutrinos have no Majorana masses. So they should naturally obtain their Dirac masses. For this purpose, we consider a $U(1)_{B-L}^{}$ gauge symmetry under which two right-handed neutrinos carry the $B-L$ number $-4$ while the standard model quarks and leptons carry the $B-L$ numbers $+1/3$ and $-1$ as usual. After a Higgs singlet spontaneously breaks the $U(1)_{B-L}^{}$ symmetry, we can simultaneously give the required lepton number violation for neutrinoless double beta decay and the tiny mass for Dirac neutrinos. The relevant diagrams are shown in Fig. \ref{dirac}. Note a new right-handed fermion with the $B-L$ number $+5$ should be necessary to cancel the $U(1)_{B-L}^{}$ gauge anomaly \cite{mp2007}. This fermion which now is massless can decouple above the QCD scale as well as the two right-handed neutrinos to safely avoid the BBN constraint if the $U(1)_{B-L}^{}$ breaking scale is high enough.

In conclusion, we have combined the original and alternative neutrinoless double beta decay processes and then have pointed out that the Majorana neutrino mass for mediating the original neutrinoless double beta decay process should include the Majorana neutrino mass induced after the alternative neutrinoless double beta decay processes. Consequently we have found that the complete Majorana neutrino mass can be allowed to arrive at a zero value by taking into account a realistic cancellation among different-sources Majorana neutrino masses. We hence can simultaneously forbid the original neutrinoless double beta decay process and realize the alternative neutrinoless double beta decay processes. As a result, even if the neutrinoless double beta decay process is fortunately observed in the future experiments, it cannot absolutely rule out the possibility of Dirac neutrinos. We should develop other experiments to eventually distinguish the Majorana and Dirac neutrinos.

When this work was finalising, Shun Zhou told us that during a meeting at the Mainz Institute for Theoretical Physics, Apostolos Pilaftsis had mentioned that the tree-level parameters could be well chosen to give a vanishing neutrino mass in the type-I seesaw model, while the original neutrinoless double beta decay process remained a nonzero rate as the nuclear medium effects on quarks broke any intricate cancellation. See the footnote in \cite{lzz2016}. Obviously Pilaftsis's unpublished scheme is totally different from ours.

\textbf{Acknowledgement:} We thank Junjie Cao, Xiao-Gang He, Yu Jia, Tianjun Li, Yu-Feng Li, Rabi N. Mohapatra, Jose W. F. Valle, Jiang-Hao Yu and Shun Zhou for helpful comments. This work was supported in part by the National Natural Science Foundation of China under Grant No. 11675100 and in part by the Fundamental Research Funds for the Central Universities.


\begin{thebibliography}{99}



\bibitem{racah1937}
G. Racah, Nuovo Cim. \textbf{14}, 322 (1937).



\bibitem{majorana1937}
E. Majorana, Nuovo Cim. \textbf{14}, 171 (1937). 




\bibitem{furry1939}
W.H. Furry, Phys. Rev. \textbf{56}, 1184 (1939).




\bibitem{fukuda1998}
Y. Fukuda {\it et al.}, Phys. Rev. Lett. \textbf{81}, 1562 (1998). 



\bibitem{fukuda2001}
Y. Fukuda {\it et al.}, Phys. Rev. Lett. \textbf{86}, 5651 (2001). 

\bibitem{ahmad2001}
Q.R. Ahmad {\it et al.}, Phys. Rev. Lett. \textbf{87}, 071301 (2001).

\bibitem{an2012}
F.P. An {\it et al.}, Phys. Rev. Lett. \textbf{108}, 171803 (2012).


\bibitem{minkowski1977}
P. Minkowski, Phys. Lett. B \textbf{67}, 421 (1977). 
  

\bibitem{yanagida1979}
T. Yanagida, {\it Proceedings of the Workshop on Unified Theory and the Baryon Number of the Universe}, 95, ed. Sawada, O. and Sugamoto, A. (Tsukuba 1979).

\bibitem{grs1979}
M. Gell-Mann, P. Ramond and R. Slansky, {\it Supergravity: Proceedings of the Supergravity Workshop at Stony Brook}, 315, ed. van Nieuwenhuizen, F. and Freedman, D. (North Holland 1979).




\bibitem{ms1980}
R.N. Mohapatra and G. Senjanovi\'{c}, Phys. Rev. Lett. \textbf{44}, 912 (1980).





\bibitem{mw1980}
M. Magg and C. Wetterich, Phys. Lett. B \textbf{94}, 61 (1980).


\bibitem{sv1980}
J. Schechter and J.W.F. Valle, Phys. Rev. D \textbf{22}, 2227 (1980).



\bibitem{cl1980}
T.P. Cheng and L.F. Li, Phys. Rev. D \textbf{22}, 2860 (1980).


\bibitem{lsw1981}
G. Lazarides, Q. Shafi and C. Wetterich, Nucl. Phys. B \textbf{181}, 287 (1981).


\bibitem{ms1981}
R.N. Mohapatra and G. Senjanovi\'{c}, Phys. Rev. D \textbf{23}, 165 (1981).



\bibitem{flhj1989}
R. Foot, H. Lew, X.G. He and G.C. Joshi, Z. Phys. C \textbf{44}, 441 (1989).




\bibitem{zee1980}
A. Zee, Phys. Lett. B \textbf{93}, 389 (1980). 


\bibitem{zee1986}
A. Zee, Nucl. Phys. B \textbf{264}, 99 (1986).

 
 
\bibitem{babu1988} 
K.S. Babu, Phys. Lett. B \textbf{203},132 (1988).

\bibitem{bl2001}
K.S. Babu and C.N. Leung, Nucl. Phys. B \textbf{619}, 667 (2001).




\bibitem{knt2003}
L.M. Krauss, S. Nasri and M. Trodden, Phys. Rev. D \textbf{67}, 085002 (2003).




\bibitem{ma2006}
E. Ma, Phys. Rev. D \textbf{73}, 077301 (2006).


\bibitem{rw1983}
M. Roncadelli and D. Wyler, Phys. Lett. B \textbf{133}, 325 (1983).


\bibitem{rs1984}
P. Roy and O. Shanker, Phys. Rev. Lett. \textbf{52}, 713 (1984).





\bibitem{sv1982}
J. Schechter and J.W.F. Valle, Phys. Rev. D \textbf{25}, 2951 (1982).







\bibitem{mohapatra1986}
R.N. Mohapatra, Phys. Rev. D \textbf{34}, 3457 (1986).

\bibitem{hkk1995}
M. Hirsch, H.V. Klapdor-Kleingrothaus and S.G. Kovalenko, Phys. Rev. Lett. \textbf{75}, 17 (1995).




\bibitem{gu2012}
P.H. Gu, Phys. Rev. D \textbf{85}, 093016 (2012).


\bibitem{bm2002}
B. Brahmacharia and E. Ma, Phys. Lett. B \textbf{536}, 259 (2002). 




\bibitem{gj2008}
A. de Gouv\^{a} and J. Jenkins, Phys. Rev. D \textbf{77}, 013008 (2008).


\bibitem{bhow2013}
F. Bonnet, M. Hirsch, T. Ota and W. Winter, JHEP \textbf{1303}, 055 (2013).



 


\bibitem{vergados2002}
J. Vergados, Phys. Rept. \textbf{361}, 1 (2002).



\bibitem{aee2008}
F.T. Avignone III, S.R. Elliott and J. Engel, Rev. Mod. Phys. \textbf{80}, 481 (2008).





\bibitem{dpr2019}
M.J. Dolinski, A.W.P. Poon and W. Rodejohann, Ann. Rev. Nucl. Part. Sci. \textbf{69}, 219 (2019).






\bibitem{mp2007}
J.C. Montero and V. Pleitez, Phys. Lett. B \textbf{675}, 64 (2009). 





\bibitem{lzz2016}
J.H. Liu, J. Zhang and S. Zhou, Phys. Lett. B \textbf{760}, 571 (2016).


\end{thebibliography}
\end{document}